\documentclass[twoside,a4paper,11pt]{sea10}
\usepackage{graphicx}
\usepackage{hyperref}
\usepackage{movie15}
\usepackage{natbib}
\newcommand\msunyr{\mbox{\,${\rm M_{\odot}\, yr^{-1}}$}}
\newcommand{\ergshz}{\mbox{~erg~s$^{-1}$~Hz$^{-1}$}}

\def\ccsnrate{\mbox{\,$r_{\rm CCSN}$}}
\newcommand{\EE}[1]{\hbox{$10^{ #1 }$}}

\newcommand{\msun}{\hbox{~${\rm M}_\odot$}}

\def\msun{\hbox{M$_{\odot}$}}

\def\msunyr{\mbox{\,${\rm M_{\odot}\, yr^{-1}}$}}

\def\degs{\ifmmode ^{\circ}\else$^{\circ}$\fi}
\def\amin{\ifmmode ^{\prime}\else$^{\prime}$\fi}
\def\asec{\ifmmode ^{\prime\prime}\else$^{\prime\prime}$\fi}
\def\arcsec{\ifmmode ^{\prime\prime}\else$^{\prime\prime}$\fi}

\def\degs{\ifmmode ^{\circ}\else$^{\circ}$\fi}
\def\amin{\ifmmode ^{\prime}\else$^{\prime}$\fi}

\def\EE#1{\times 10^{#1}}

\def\cm{\mbox{\,cm}}

\def\cm3{\mbox{\,cm$^{-3}$}}

\def\ergshz{\mbox{~erg~s$^{-1}$~Hz$^{-1}$}}

\def\lsim{\!\!\!\phantom{\le}\smash{\buildrel{}\over
 {\lower2.5dd\hbox{$\buildrel{\lower2dd\hbox{$\displaystyle<$}}\over
                                 \sim$}}}\,\,}
\def\gsim{\!\!\!\phantom{\ge}\smash{\buildrel{}\over
{\lower2.5dd\hbox{$\buildrel{\lower2dd\hbox{$\displaystyle>$}}\over
                               \sim$}}}\,\,}
\def\aj{AJ}%
\def\araa{ARA\&A}%
\def\apj{ApJ}%
\def\aap{A\&A}%
\def\mnras{MNRAS}%
\begin{document}
\pagenumbering{arabic}
\pagestyle{myheadings}
\thispagestyle{empty}
{\flushleft\includegraphics[width=\textwidth,bb=58 650 590 680]{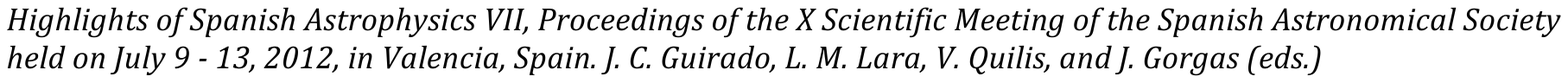}}
\vspace*{0.2cm}
\begin{flushleft}
{\bf {\LARGE
%
Digging deep into the ULIRG phenomenon: When radio beats dust.
}\\
\vspace*{1cm}
%
Miguel \'Angel P\'erez-Torres$^{1}$
%
}\\
\vspace*{0.5cm}
%
$^{1}$
IAA-CSIC, Glorieta de la Astronom\'ia s/n, 18008 Granada, Spain\\
%
\end{flushleft}
%
\markboth{
Digging deep into the ULIRG phenomenon: When radio beats dust.
}{ 
%
Miguel \'Angel P\'erez-Torres
%
}
\thispagestyle{empty}
\vspace*{0.4cm}
\begin{minipage}[l]{0.09\textwidth}
\ 
\end{minipage}
\begin{minipage}[r]{0.9\textwidth}
\vspace{1cm}
\section*{Abstract}{\small
%
Luminous and Ultra-Luminous Infrared Galaxies (U/LIRGs) do also radiate copious amounts of radio emission, both thermal (free-free) and non-thermal (mainly synchrotron). This is very handy since, unlike optical and infra-red observations, radio is not obscured by the ubiquitous dust present in  U/LIRGs, which allows a direct view of the ongoing activity in the hearts of those prolific star-forming galaxies.
Here, I first justify the need for this high-angular resolution radio studies of local U/LIRGs, discuss the energy budget and the magnetic field, as well as IC and synchrotron losses in U/LIRGs, and present some selected results obtained by our team on high-angular resolution radio continuum studies of U/LIRGs.
Among other results, I show the impressive discovery of an extremely prolific supernova factory in the central $\sim$150 pc of the galaxy Arp 299-A (D=45 Mpc)  and the monitoring of a large number of very compact radio sources in it, the detection and precise location of the long-sought AGN in Arp 299-A. A movie summarizing those results can be found in 
\url{http://www.iaa.es/~torres/research/arp299a.html}.
  All those results demonstrate that very-high angular resolution studies of nearby U/LIRGs are of high relevance for the comprehension of both local and high-z starbursting galaxies.

%
\normalsize}
\end{minipage}
%
%
%
\section{Introduction \label{intro}}

The study of galaxies at the highest infrared luminosities, i.e., Luminous 
$10^{11}\, {\rm L_\odot} \leq {\rm L_{\rm IR}} \le 10^{12}\, {\rm L_\odot}$)
and Ultra-Luminous  ($L{_{\rm IR}} \geq 10^{12}\, L{_\odot}$) Infrared Galaxies (LIRGs and ULIRGs, respectively) is relevant for many reasons:

\begin{itemize}
\item U/LIRGs account for about 10\% of the total radiant energy production and $\sim$ 20\% of all the high-mass star-formation in the local universe (e.g., \citealt{brinchmann04}). 
\item The bulk of the energy radiated by ULIRGs is infrared emission from warm dust
grains heated by a central power source, or sources, whose nature (AGN, starburst, or both) is often unknown, but whose evolution is likely to be related.
\item Essentially all LIRGs above $\sim 10^{11.4}\,  {\rm L_\odot}$ appear to be interacting galaxies, and all ULIRGs seem to be advanced merger systems. Thus they may represent an important stage in the formation of quasi-stellar objects.  
\item U/LIRGs are also of cosmological relevance, as they bear many similarities with star-forming galaxies at high-$z$ (e.g.
local UV-bright starbursts are good analogs to Lyman Break Galaxies (e.g. \citealt{meurer97}). 
 \item Finally, U/LIRGs are ideal laboratories to study the complex ecosystems of stars, gas, Black Holes, the interstellar medium (ISM), and test massive star and black-hole/starburst evolution, or probe the Initial Mass Function (IMF) of the high-mass stars.
\end{itemize}

\section{The dust heating central source: AGN vs. SB}

 The critical
question concerning U/LIRGs is whether the dust in the central
regions ($r \lsim$1~kpc) is heated by a starburst or an active
galactic nucleus (AGN), or a combination of both. Mid-infrared
spectroscopic studies on samples of ULIRGs 
suggested that the vast majority of these galaxies were powered
predominantly by recently formed massive stars, with a significant
heating from the AGN only in the most luminous objects 
\citep{veilleux99}.  \citeauthor{veilleux99} also found that at least half
of ULIRGs are probably powered by both an AGN and a starburst in a
circumnuclear disk or ring.  These circumnuclear regions are typically located
at radii $r\lsim$700~pc from the nucleus of the galaxy, and also
contain large quantities of dust. Those facts (compactness of the emission and dusty environments) prevent to gain much insight into the nature of the central source, either from optical, or infrared observations.


\emph{Why radio observations?}
The largely dusty environments of U/LIRGs imply huge extinctions in the optical, and even in the infrared, which prevents the direct detection of recent starburst activity, e.g., via supernova explosions. The advantage of observing in radio is that these wavelengths are not affected by dust obscuration, thus allowing to see the activity within those deeply buried regions.

\emph{Why high-angular resolution?}
At a typical distance of 100 Mpc to a local U/LIRG, 1'' corresponds to a linear size of about 500 pc. Therefore, if one aims at spatially resolving the circumnuclear regions of U/LIRGs to understand in detail the ongoing processes in the central regions of those monsters, sub-arcsecond resolution is a must. This makes an excellent case for high-angular resolution radio observations of U/LIRGs, since current radio interferometric arrays provide sub-arcsecond angular resolution (EVLA, MERLIN) or even milliarsecond resolution, using Very Long Baseline Interferometry (VLBI).

\emph{How do U/LIRGs look in radio?}
The central kiloparsec region of many nearby U/LIRGs shows a distribution of their radio emission typically consisting of a compact ($\lsim$ 200 pc), high surface brightness, central radio source corresponding to an active nuclear region (whether powered by an AGN, or a nuclear starburst), immersed in a low surface brightness circumnuclear, extended ($\simeq$ 1 kpc) halo \citep{condon91}. 

\begin{figure}
\center
\includegraphics[width=\textwidth]{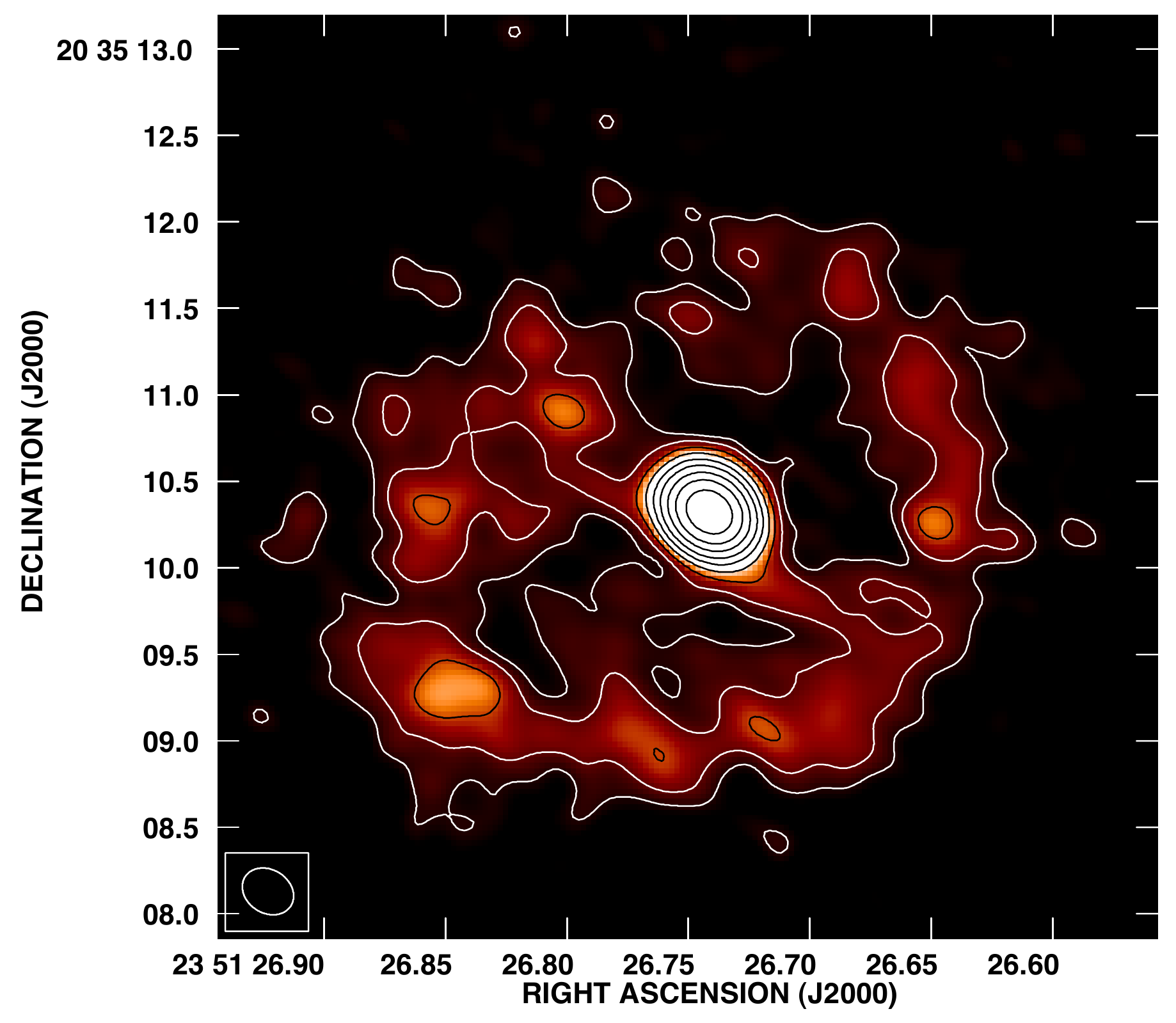}
\caption{
\label{fig1}
The LIRG Mrk~331 (D=75~Mpc) imaged at 8.46 GHz with the VLA on December 2006 (unpublished image). 
The synthesized FWHM beam is a Gaussian beam of size 0.27''$\times$ 0.25''. Contour levels are drawn at
  (3,$3\,\sqrt{3}$,9,...)$\times 19\mu$ Jy beam$^{-1}$, the
off-source rms of the image, and the peak of brightness is 7.3~mJy beam$^{-1}$, and corresponds to the bright spot 
in the central region of the galaxy. 1\asec corresponds to 357 pc. (See text for details.)
}
\end{figure}

As an example, I show in Figure \ref{fig1} an 8.46 GHz VLA image of the LIRG Mrk 331. At a distance of D=75~Mpc to Mrk~331, 1\arcsec\ corresponds to a linear size of 357 pc. Note that all radio emission is concentrated in a region whose angular size is less that 3\arcsec\ in diameter, or less than 1~kpc. 
The total 8.5 GHz flux density is of $\simeq$ 29.3~mJy, corresponding to a monochromatic radio luminosity of $
\sim 1.6\EE{29} \ergshz$.

\begin{figure}
\center
\includegraphics[width=\textwidth]{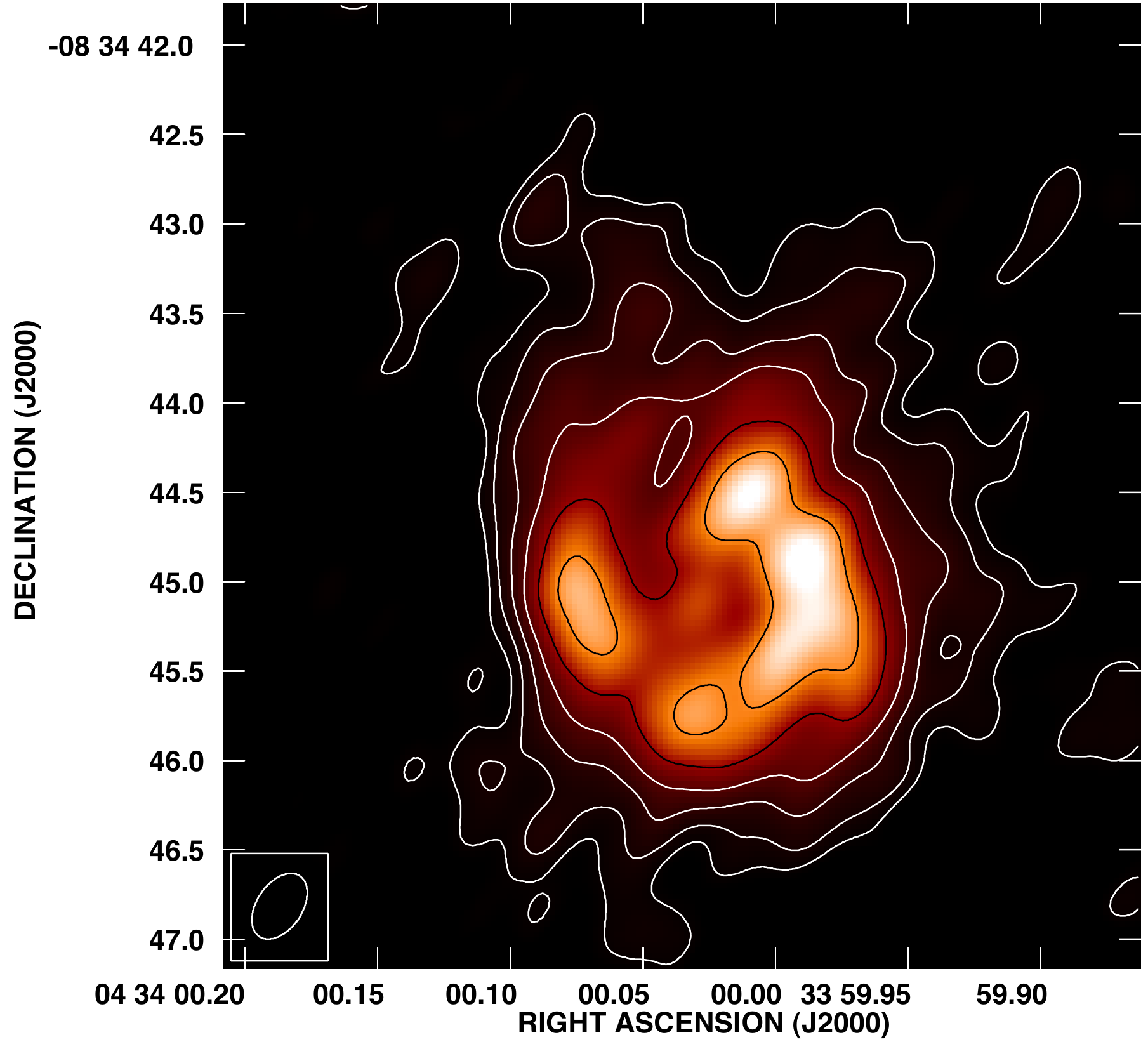}
\caption{
\label{fig2}
The LIRG NGC~1614 (D=65.6~Mpc) imaged at 8.46 GHz with the VLA on June 2006 (unpublished image). 
The synthesized FWHM beam is a Gaussian beam of size 0.42''$\times$ 0.24''. Contour levels are drawn at
  (3,$3\,\sqrt{3}$,9,...)$\times 22\mu$ Jy beam$^{-1}$, the
off-source rms of the image, and the peak of brightness is 1.78~mJy beam$^{-1}$, and corresponds to a bright spot 
in the circumnuclear ring of star-formation. 1\asec corresponds to 320 pc. (See text for details.)
}
\end{figure}

Often, the high-surface brightness is not within the central $\sim$200 pc region, but on a prominent circumnuclear ring where a burst of recent star formation is ongoing.
Figure \ref{fig2} shows an 8.46 GHz VLA image of the LIRG NGC~1614. At the distance of D=65.6~Mpc to NGC~1614, 1\arcsec\ corresponds to a linear size of 320 pc. 
As with Mrk 331, all radio emission is concentrated in a region less than 1~kpc in size. 
The total 8.5 GHz flux density is of $\simeq$ 32.9~mJy, corresponding to a monochromatic radio luminosity of $
\sim 1.8\EE{29} \ergshz$, and I note that unlike the case of Mrk 331, where the circumnuclear ring is very faint, about 90\% of the radio luminosity from NGC 1614 comes from a prominent circumnuclear ring which is at a distance of $\sim$250 pc from the nucleus of the galaxy.

Based on those high-angular radio observations, it is evident that, contrary to what we could have thought from a look at a low-angular resolution image, most of the starburst activity happening in NGC~1614 is actually taking place in this ring. By converting the measured flux density into a brightness temperature, $T_{\rm b}$, we would find out that this temperature is, in most regions of the ring, much higher than the thermal temperature of the plasma around the massive stars, implying a non-thermal (synchrotron) origin for the radio emission. The natural explanation for this non-thermal emission is that it is due to recent explosions of core-collapse supernovae (CCSNe) and supernova remnants (SNRs), which implies that most of the starburst activity is still ongoing ($t \lsim 40$ Myr) in the ring of NGC~1614.

\section{Radio as a direct tracer of massive star-formation in U/LIRGs}
\label{radio:sf}

A large fraction of the massive star-formation at both low- and
high-$z$ has taken place in U/LIRGs \citep{magnelli11}. 
Indeed, by applying evolutionary stellar models to bursts of age 10-100 Myr, and adopting a standard IMF, one can obtain the SFR as a function of $L_{\rm FIR}$ \citep{kennicutt98}:

\begin{equation} \label{eq:sfr}
{\rm SFR} \approx 17.2 \left( \frac{\rm L_{\rm FIR}}{10^{11}\,{\rm L_\odot}}\right) \, {\rm M_\odot \,  yr^{-1}} 
\end{equation}

While the compact, centrally located radio emission in these galaxies might be generated by a point-like source (AGN), or by the combined effect of multiple RSNe and supernova remnants (SNRs), e.g., the evolved nuclear starburst in Arp~220 (Parra et al. 2007 and references therein) and M~82 (e.g. Muxlow et al. 1994), it seems now well established that in the circumnuclear regions of those objects there is an ongoing burst of star formation producing core-collapse supernovae (CCSNe, Type Ib/c and Type II SNe) at a high rate (e.g.  NGC 7469, Colina et al. 2001a,b; Arp 299, Neff et al. 2004; NGC 6240, Gallimore \& Beswick 2004; Arp 220, Parra et al. 2007; IRAS~18293-3413, Mattila et al. 2007, P\'erez-Torres et al. 2007; IRAS 17138-1017, Kankare et al. 2008a, P\'erez-Torres et al. 2008, Kankare et al. 2008b).
Therefore, a powerful tracer for (recent) starburst activity in
U/LIRGs is the detection of CCSNe.

Since  stars more massive than M $\gsim$8 M$_{\odot}$ explode as CCSNe, the observed rate at which those stars die (\ccsnrate)  can be used as a direct measurement of the current star formation rate (SFR) in galaxies, and may provide unique  information about the initial mass function (IMF) of the high-mass stars. Indeed, if for the sake of simplicity we assume a constant star-formation rate for a starburst, it can be shown that (e.g. \citealt{pereztorres09b})

\begin{equation} \label{eq:ccsnrate}
\ccsnrate  =  SFR\, \left( \frac{\alpha-2}{\alpha-1} \right) \left(
          \frac{m_{\rm SN}^{1-\alpha} - m_u^{1-\alpha}}{m_l^{2-\alpha} 
           - m_u^{2-\alpha}} \right)
\end{equation}

\noindent
where $SFR$ is the constant star formation rate in \msunyr, $m_l$ and $m_u$ are the lower and upper mass limits of the initial mass function (IMF, $\Phi \propto m^{-\alpha}$), and $m_{\rm SN}$ is the minimum mass of stars that yield supernovae, assumed to be 8 \msun (e.g., \citealt{smartt09}).  

  While the rate at which stars die in normal galaxies is rather low (e.g., one SN is expected to explode in the Milky Way every $\sim$50 yr, ), the CCSN rate in U/LIRGs is expected to be at least one to two orders of magnitude higher than in normal galaxies, and hence detections of SNe in U/LIRGs offer a promising way of determining the current star formation rate in nearby galaxies. In fact, from Equation \ref{eq:sfr}, and assuming 
a Salpeter IMF ($\alpha=2.35$) with 
$m_u$=100 \msun and $m_l$=0.2 \msun, we obtain that a LIRG (ULIRG) is expected to yield about 0.26 (2.6) SN/yr, or more.
Thus, equations \ref{eq:sfr} and \ref{eq:ccsnrate} imply that U/LIRGs are prolific star-forming factories, with LIRGs having typical values of SFRs and core-collapse supernova rates of $\sim$(17-170) \msunyr and $\ccsnrate \sim$(0.3-3.0) yr$^{-1}$, respectively, with ULIRGs exceeding those upper ends.



However, the direct detection of CCSNe in the extreme densities of the central few hundred pc of U/LIRGs is, as pointed out earlier, extremely difficult since emission in the visual band suffers very significant extinction in those regions, which contain large amounts of dust, and can at best yield only a lower limit to the true value of \ccsnrate.  
Fortunately, radio is not affected by dust absorption. Moreover, significant radio emission from CCSNe is expected --as opposed to thermonuclear SNe-- as the interaction of the SN ejecta with the circumstellar medium (CSM) gives rise to a high-energy density shell, which is Rayleigh-Taylor unstable and drives turbulent motions that amplify the existing magnetic field in the presupernova wind, and efficiently accelerate relativistic electrons, thus enhancing the emission of synchrotron radiation at radio wavelengths \citep{chevalier82}. 

Thus, starburst activity in the circumnuclear regions of LIRGs ensures both the presence of a high number of massive stars and a dense surrounding medium, so bright radio SNe are expected to occur. Those SNe can be detected, even in the innermost regions of U/LIRGs, by means of high angular resolution ($\leq 0.05$ arcsec), high-sensitivity ($\leq$0.05 mJy) radio searches of CCSNe, as radio emission is unaffected by dust extinction, and the angular resolution yielded by current Very Long Baseline Interferometry (VLBI) arrays, of the order of a few milliarcsec at cm-wavelengths,  is able to unveil nuclear starbursts and detect individual radio supernovae at large distances in the local Universe (e.g. Arp~220, \citealt{parra07};Arp 299, \citealt{pereztorres09b,bondi12};  see also Figure \ref{fig3}; IC~883 \citealt{romero-canizales12b}).

\section{The local luminous infra-red galaxy Arp~299 as a case study}

\subsection{An extremely prolific supernova factory in Arp 299}

\begin{figure}
\center
\includegraphics[scale=0.39]{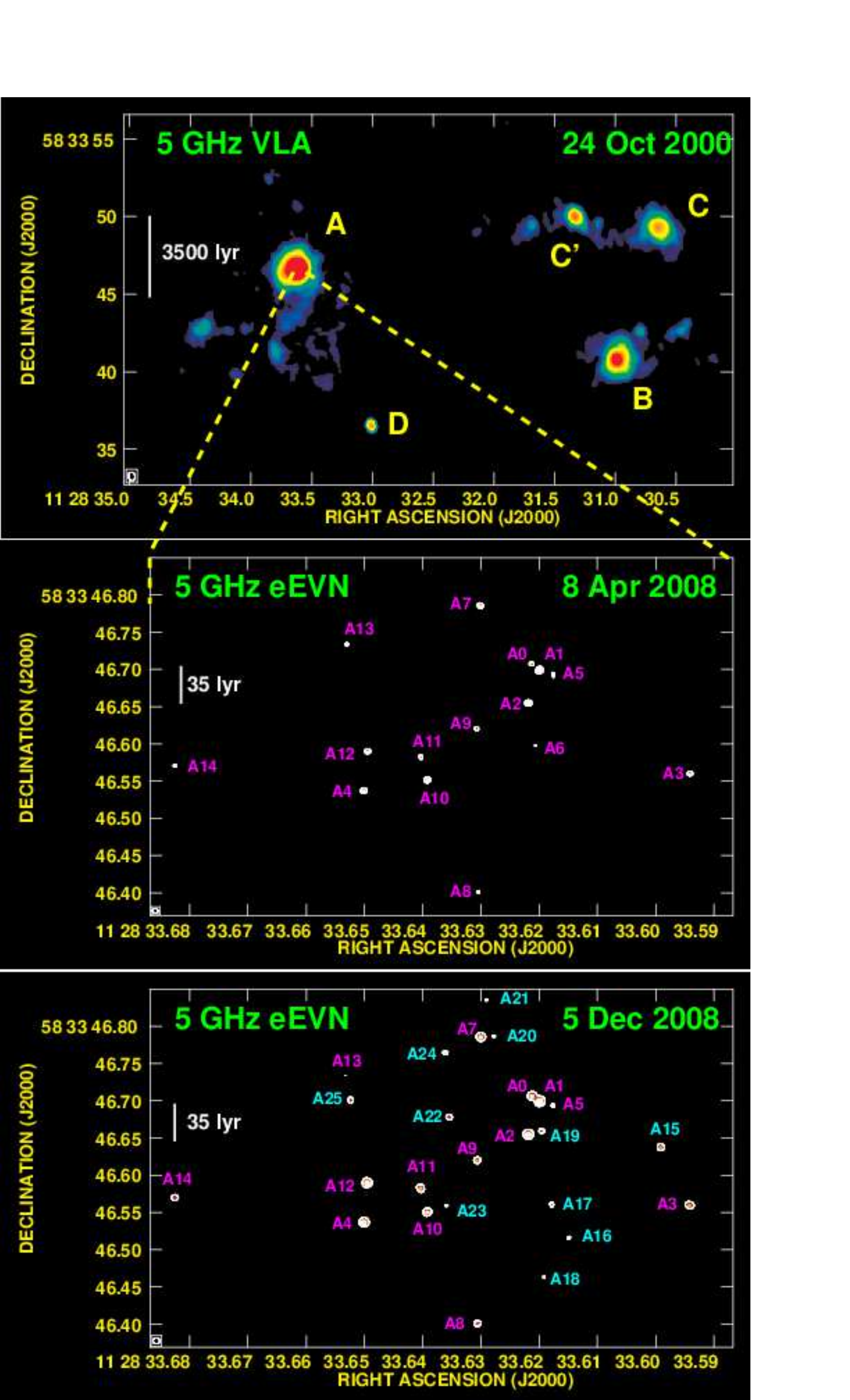}
\includegraphics[scale=0.355]{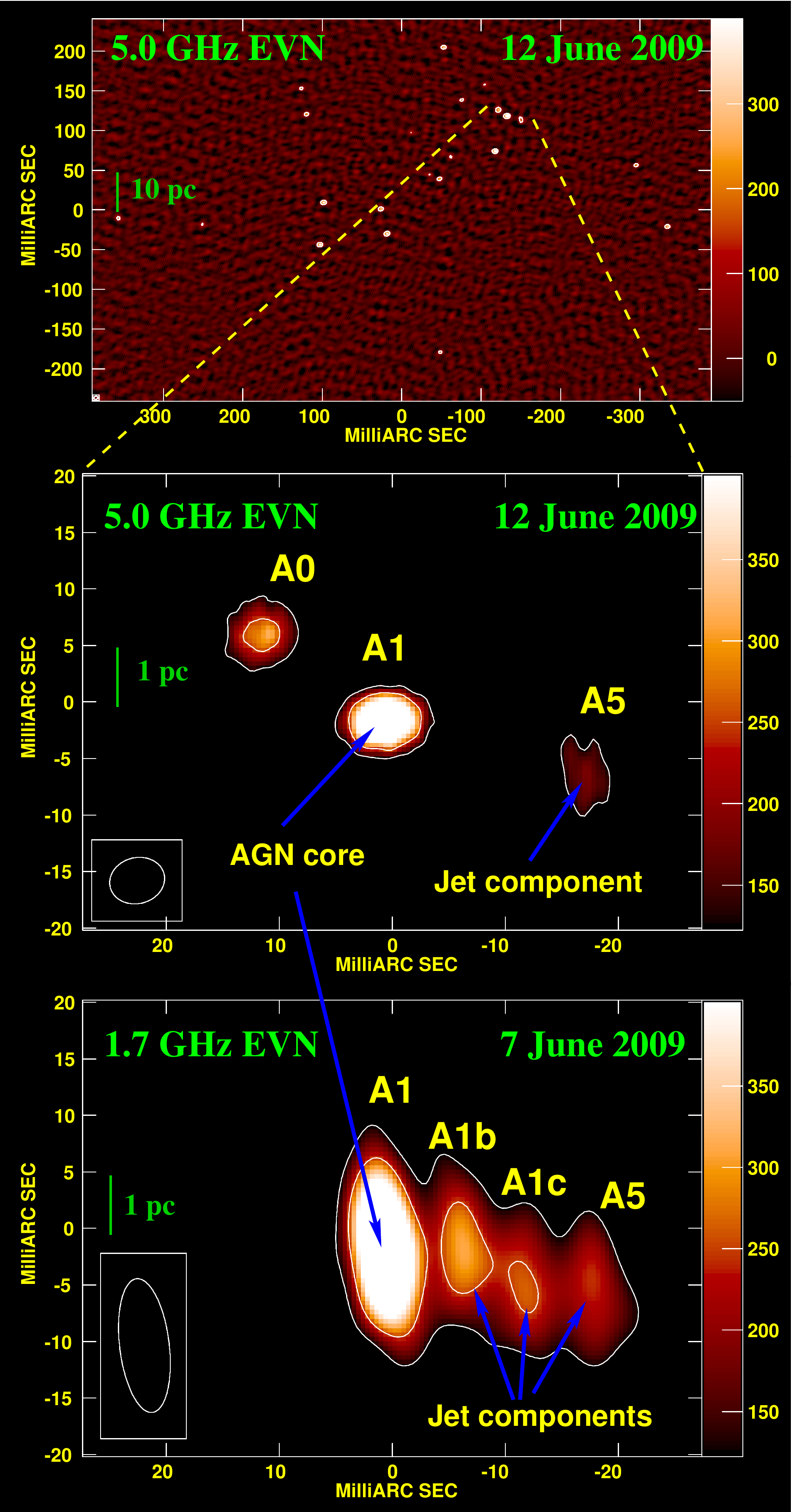} 
\caption{ \label{fig3}
\small {\it
  Top left:} 5 GHz VLA archival observations of Arp 299 on 24 October 2000,
displaying the five brightest knots of radio emission in this merging
galaxy. 
{\it  Middle left and bottom left:} Contour maps drawn at five times the r.m.s. of our 5 GHz eEVN observations of the central 500 light
years of the luminous infrared galaxy Arp 299-A on 8 April 2008 and 5
 December 2008, revealing a large population of relatively bright, compact, non-thermal emitting sources. The size of
 the FWHM synthesized interferometric beam was (0.6 arcsec $\times$
 0.4 arcsec) for the VLA observations, and (7.3 milliarcsec $\times$
 6.3 milliarcsec) and (8.6 milliarcsec $\times$ 8.4 milliarcsec) for
 the EVN observations on 8 April 2008 and 5 December 2008,
 respectively.  To guide the reader's eye, we show in cyan the components
 detected only at the 5 December 2008 epoch.
 {\it Top right:} 5.0 GHz full EVN image of the central 150 parsec
  region of the luminous infrared galaxy Arp 299-A (=IC 694), 
displaying a large number of bright, compact, nonthermal
  emitting sources, mostly identified with young RSNe and
 SNRs. The image center is at RA 11:28:33.63686 and
 DEC 58:33:46.5806.  
 {\it Middle left and bottom left:} Blow-ups of the inner
 8 parsec of the nuclear region of Arp 299-A, as imaged with the full
 EVN at 1.7 and 5.0 GHz.  The image center is at RA 11:28:33.61984
 and DEC 58:33:46.7006 in both panels. The morphology, spectral
 index and luminosity of the A1-A5 region are very suggestive of a
 core-jet structure. Contours are drawn at 5 and 10 times
 the off-source r.m.s. noise.
 }
\end{figure}

Arp~299 consists of two interacting galaxies (IC 694 and NGC 3690), which are in an early merger stage. At a luminosity distance of 44.8 Mpc \citep{fixsen96} for $H_0 = 73$~km~s$^{-1}$~Mpc$^{-1}$, Arp~299 has an infrared luminosity $L_{\rm IR} \approx 6.7\EE{11} L_{\odot}$, which almost qualifies it as a ULIRG.  The innermost $\sim$150 pc nuclear region of Arp 299-A (see top left panel in Fig. \ref{fig3}) is heavily dust-enshrouded,  thus making the detections of SNe very challenging even at near-infrared wavelengths. Yet, Arp 299 hosts recent and intense star-forming activity, as indicated by the relatively high frequency of {supernovae discovered at optical and near-infrared wavelengths} in its outer, much less extinguished regions.

The brightest component at infrared and radio wavelengths is IC 694 (A in the top panel of Fig. \ref{fig3}; hereafter Arp 299-A), which accounts for $\sim$50\% of the total infrared luminosity of the system \citep{almudena00}, and $\sim$70\%\/ of its 5 GHz radio emission \citep{neff04}. Numerous H~II regions populate the system near star-forming regions, which implies that star formation has been occurring at a high rate for past $\sim$10 Myr \citep{almudena00}.  Given that IC~694 accounts for most of the infrared emission in Arp 299, it is the region that is most likely to contain new SNe \citep{condon92}. Since optical and near-infrared observations are likely to miss a significant fraction of CCSNe in the innermost regions of Arp 299-A due to  high values of extinction [$A_V \sim 34-40$ \citep{almudena09}] and the lack of the necessary angular resolution, radio observations of Arp 299-A at high angular resolution, high sensitivity are the only way of detecting new CCSNe and measuring directly and independently of any model its CCSN and star formation rates.  Very Long Baseline Array (VLBA) observations carried out during 2002 and 2003 resulted in the detection of five compact sources \citep{neff04}, one of which (A0) was identified as a young SN.

\subsection{The AGN-Starburst connection in Arp 299-A}
\label{agn-sb}

As mentioned in the introduction of this contribution, the dusty nuclear regions of luminous infra-red galaxies (LIRGs) are
heated by either an intense burst of massive star formation, or an
  active galactic nucleus (AGN), or a combination of both.
  Disentangling the contribution of each of those putative dust-heating agents is a challenging task, and direct imaging of the   innermost few pc can only be accomplished at radio wavelengths, 
  using very high-angular  resolution observations.

My team and I started to monitor the A-nucleus of the interacting starburst galaxy
  Arp 299, using  very long baseline interferometry (VLBI)
  radio observations at 1.7 and 5.0 GHz \citep{pereztorres09a}.  Our aim was to characterize
  the compact sources in the innermost few pc region of Arp
  299-A, as well as to detect recently exploded core-collapse
  supernovae.  
We imaged with milliarcsecond resolution the inner eight parsecs of the
nuclear region in Arp 299-A, using contemporaneous EVN observations at
1.7 and 5.0 GHz \citep{pereztorres10}, and the summary of those results is shown in the right panel of Fig. \ref{fig3}.
 At 5.0 GHz, we detect components A0, A1, and A5, as
previously detected by \citet{pereztorres09a}. At 1.7 GHz, A1 and A5
are also detected, but not A0. In addition, two new components  between A1 and
A5 (A1b and A1c) are detected at 1.7 GHz, forming a complex.

The morphology, spectral index, radio luminosity, and
radio-to-X ray luminosity ratio of the A1-A5 complex are consistent
with that of an LLAGN, and rules out the possibility that it consists
of a chain of young RSNe and SNRs in a young SSC. We therefore
concluded that A1 is the long-sought AGN in Arp 299-A.
Since Arp 299-A had long been thought of as a pure starburst, our finding of a buried, low-luminosity AGN in its central region, coexisting with a recent burst of starformation, suggests that both a starburst and AGN are frequently associated phenomena in mergers. In this case, our result is likely to have an impact on  
evolutionary scenarios proposed for AGN and the triggering mechanism of activity in general.

We note that component A0, previously identified as a
young RSN by \citet{neff04}, is not seen at our low-frequency observations,
which implies there must be a foreground absorbing H~II region, which precludes its detections at frequencies $\lsim$1.7 GHz.  It is remarkable that this RSN exploded at the  mere (deprojected) distance
of two parsecs from the putative AGN in Arp 299-A, which
makes this supernova one of the closest to a central supermassive
black hole ever detected. This result may also be relevant to
accreting models in the central regions of galaxies, since it is not
easy to explain the existence of very massive, supernova progenitor
stars so close to an AGN. While seemingly contradictory, this could
explain the low-luminosity of the AGN we see in Arp 299-A. In
fact, since massive stars shed large amounts of mechanical energy into
their surrounding medium, thereby significantly increasing its
temperature, those massive stars would hinder the accretion of material to the
central black hole, which could in turn result in a less powerful AGN
than usual. (See also \citet{herrero12} and the contribution of Herrero-Illana et al. to these proceedings.)

\section{Energy budget in U/LIRGs}
\label{energy}

Since essentially all of the observed radio emission in the central regions of U/LIRGs is of synchrotron origin
(see section 2.1), we can estimate a minimum total energy in magnetic fields,
electrons, and heavy particles and a minimum magnetic field for them. In fact, if we assume equipartition (magnetic field energy is approximately equal 
to the total particle energy), then  
the minimum total energy is \citep{pacholczyk70}

\begin{equation}
E_{\rm min}^{\rm (tot)} = c_{13}\, (1 + k)^{4/7}\, \phi^{3/7}\, R^{9/7}\, L^{4/7}
\end{equation}

\noindent
where $L$ is the radio luminosity of the source between $\nu_1$ and $\nu_2$,
$R$ is a characteristic size,
$c_{13}$ is a slowly-dependent function of the observed radio spectral index, $\alpha$, 
$\phi$ is the fraction of the supernova's volume occupied  by the magnetic field and 
by the relativistic particles (filling factor), and
$k$ is the ratio of the (total) heavy particle energy to the electron energy.
This ratio depends on the mechanism that generates the relativistic electrons, ranging
from $k \approx 1$ up to $k = m_p/m_ e \approx 2000$, where $m_p$ and $m_e$  
are the proton and electron mass, respectively. A value of $k \simeq 100$ seems appropriate for most galaxies \citep{beck05}.
For the sake of simplicity, I will use a value of $\alpha = -0.7$, which is a typical radio synchrotron spectral index.
The value of the function $c_{13}$ is  $\sim 2.4 \times 10^4$
(for $\alpha =-0.7$, $\nu_1=30$ MHz and $\nu_2=$30 GHz). 
The characteristic size of the synchrotron emitting region in U/LIRGs varies from case to case, although the most compact starbursts, typically residing in ULIRGs, i.e., advanced mergers, have sizes of $\sim$100-200 pc, and for the sake of simplicity we will use here a nominal value of $R=100$ pc, 
and  a filling factor value, $\phi=0.5$.
The total (synchrotron) radio luminosity between  $\nu_1=30$ MHz and $\nu_2=$30 GHz (for $\alpha = -0.7$) can be written as 

\begin{equation}
L_{\rm R} \approx 5\EE{39} \, \left( \frac{S_{\nu,0}}{10\, \rm mJy} \right) \, \left(\frac{D}{100\, \rm Mpc }\right)^2 \,{\rm erg\,s^{-1}}
\end{equation}

\noindent
where $S_{\nu,0}$ is the flux density, in mJy, at a reference frequency.
With these values, the minimum total energy --and its associated magnetic field--enclosed in a 100 pc region of a U/LIRG 
at a nominal  distance of 100 Mpc with $S_{\nu,0}$=10 mJy are

\begin{equation}
 E_{\rm min}^{\rm (tot)} \approx 1.9\EE{53} (1+k)^{4/7}\,  \left( \frac{S_{\nu,0}}{10\, \rm mJy} \right)^{4/7} \, \left(\frac{\phi}{0.5}\right)^{3/7}\, \left(\frac{R}{100\, {\rm pc}}\right)^{9/7}\,
   \left(\frac{D}{100\, {\rm Mpc}}\right)^{14/7}\,  {\rm erg}
\end{equation}

\begin{equation}
 B_{\rm min} \approx 128\,
 (1+k)^{2/7}  \left( \frac{S_{\nu,0}}{10\, \rm mJy} \right)^{2/7} \, \left(\frac{\phi}{0.5}\right)^{-2/7}\, \left(\frac{R}{100\, {\rm pc}}\right)^{-6/7}\,
\left(\frac{D}{100\, {\rm Mpc}}\right)^{4/7}\,
   {\mu \rm G}
\end{equation}

 
\noindent
where we have set $c_{12} = 10^7$ in the above equation, for simplicity.
Since $1 \leq k \leq 2000$,  $E_{\rm min}^{\rm (tot)}$ is then between $2.8\EE{53}$ erg and 
$1.5\EE{55}$ erg, and equals $2.7\EE{54}$ erg for our nominal value of $k=100$.
The corresponding values of the equipartition magnetic field lie then  between $\sim 156 \mu$G (electrons dominate)  and
$\sim 1.1\, {\rm mG}$ (heavy particles dominate), and equals $\sim 0.5$ mG for our nominal value of $k=100$.
We see that the energy contained in the central, compact regions (size $\lsim$ 100 pc) of U/LIRGs are huge, and the magnetic fields in those ambients are expected to be significant, since the above estimates are lower limits to the real values of the magnetic fields. Values of several hundreds of $\mu$G, or larger, have been obtained for the central regions of the ULIRGs IRAS~23365+3604 \citep{romero-canizales12a} and IRAS 17208-0014 \citep{momjian03}, in agreement with the advanced-merger scenarios proposed for those ULIRGs.


\section{Magnetic field and radiation field in U/LIRGs}
\label{bfield.2}

We have just seen that the minimum equipartition total energy in the form of particles and magnetic field emitting by synchrotron radio emission within the central $\sim$100 pc of a U/LIRG is huge. In particular, 
the magnetic field is expected to be significantly larger than in normal galaxies, where on average has a value of $\sim 5 \mu$G, and hence is expected to play a significant role in the physics of those objects. Similarly, since the bolometric luminosity of those beasts is dominated by radiative emission at IR wavelengths, it is also expected that the radiative field plays a significant role. {\it But which is the main player here:  the radio synchrotron, or the radiative field?}
In the previous section, we obtained that a typical magnetic field for the central $\sim$100 pc of a U/LIRG is $\sim 500 \mu$G (assuming $k\approx 100$). Therefore, we can write the energy density of the magnetic field as 

\begin{equation}
U_{\rm B} = B^2 / 8 \pi \approx  1.0 \times 10^{-8} \left(\frac{B}{500 \mu \rm G}\right)^2 {\rm erg\, cm^{-3}}
\end{equation}

\noindent
Since the bolometric luminosity of U/LIRGs is dominated by the emission at infrared wavelengths, the energy density of the radiation field can then be written as

\begin{equation}
U_{\rm rad} \approx \frac{\rm
    L_{ir}}{c\,R^2} \approx 1.3 \times 10^{-7}\,\left( \frac{L_{\rm ir}}{10^{11}\, L_\odot}\right) \, \left( \frac{R}{100 \, \rm pc}\right)^{-2}\, {\rm erg\, cm^{-3}}
\end{equation}    

Thus, {\it IC losses are likely to dominate the central $\sim$100 pc regions of U/LIRGs, if magnetic fields are close to minimum equipartition estimates}. We note, though, that it may well be that equipartition does not hold in those compact regions. For example,  \citet{thompson06} suggest that minimum equipartition estimates are not reliable, if the nuclear disks in the central regions of U/LIRGs are magnetic field supported. In the former case, i.e., when the pressure of the magnetic field is large enough to compete with that of the gas in the nuclear disk, the implied magnetic fields would be high enough so that synchrotron losses would dominate over IC losses.

\section{Summary}

U/LIRGs are embedded in extremely dusty environments, which imply huge extinctions in the optical, and even in the infrared, thus preventing the direct detection of recent starburst activity, e.g., via supernova explosions. Radio is not affected by dust obscuration, thus allowing to see the activity within those deeply buried regions. In addition, at the typical distances of $\sim$100 Mpc to a local U/LIRG, sub-arcsecond angular resolution is mandatory to spatially resolve the circumnuclear regions of U/LIRGs, so as to understand in detail the ongoing processes in their central regions.  Current radio interferometric arrays provide sub-arcsecond angular resolution (EVLA, MERLIN) or even milliarsecond resolution, using Very Long Baseline Interferometry (VLBI), and thus they become a very powerful tool to study U/LIRGs.

Here, I have presented some of the results obtained by our team, making use of high-angular resolution radio observations, including the impressive discovery of an extremely prolific supernova factory in the central $\sim$150 pc of the galaxy Arp 299-A (D=45 Mpc)  and the monitoring of a large number of very compact radio sources in it, or the detection and precise location of the long-sought AGN in Arp 299-A. In addition, VLBI observations of distant ULIRGs in the local universe have allowed us to estimate the energetics in magnetic fields and particles. The magnetic field estimates, of a few hundred $\mu$G within the $\sim$200 pc or so, are in agreement with evolutionary scenarios that suggest those ULIRGs are advanced mergers. More recently, we also used VLBI observations of to show evidence for the existence of nuclear disks ($\lsim$100 pc in size) in starburst galaxies from their radial distribution of supernovae. All those results show that very-high angular resolution studies of nearby U/LIRGs are of high relevance for the comprehension of both local and high-z starforming galaxies.

%
%
\small  
%
%

%

%

%
%
%

\end{document}